\documentclass[sn-mathphys,Numbered]{sn-jnl}
\usepackage{graphicx}%
\usepackage{multirow}%
\usepackage{amsmath,amssymb,amsfonts}%
\usepackage{amsthm}%
\usepackage{mathrsfs}%
\usepackage[title]{appendix}%
\usepackage{xcolor}%
\usepackage{textcomp}%
\usepackage{manyfoot}%
\usepackage{booktabs}%
\usepackage{algorithm}%
\usepackage{algorithmicx}%
\usepackage{algpseudocode}%
\usepackage{listings}%

\begin{document}

\title[Article Title]{Theory of gravitational lensing on a curved cosmic string}


\author*[1,2]{\fnm{Igor I.} \sur{Bulygin}}\email{8.2bulygin@gmail.com}

\author[1]{\fnm{Mikhail V. } \sur{Sazhin}}\email{moimaitre@mail.ru}
\equalcont{These authors contributed equally to this work.}

\author[1]{\fnm{Olga S.} \sur{Sazhina}}\email{cosmologia@yandex.ru}
\equalcont{These authors contributed equally to this work.}

\affil*[1]{\orgname{Sternberg Astronomical Institute of Lomonosov Moscow State University}, \orgaddress{\street{Universitetsky pr., 13}, \city{Moscow}, \postcode{119234}, \country{RF}}}

\affil*[2]{\orgname{Astrophysical school ``Traektoria''}, \orgaddress{\city{Moscow}, \postcode{107078}, \country{RF}}}


\abstract{It is discussed in detail the complete mathematical model of gravitational lensing on a single cosmic string (CS) of general shape and position with respect to the line of sight.
CS are one-dimensional extended objects assuredly predicted by modern cosmology. The presence of CS changes the global geometry of the Universe, could clarify the properties of the early Universe, including inflation models, and could serve as a unique proof of higher-dimensional theories. 
Despite the fact that CS have not yet been reliable detected, there are several strong independent indications of the existence of the CS, based of CMB analysis and search of gravitational lens chains with special properties. However, early considered models of straight CS presented only a small fraction of the general CS-configurations to be observed.
Now we propose model which could significantly increase the possibilities of CS observational search. It is considered more realistic models have necessarily include the inclinations and bends of the CS.  Besides, the recent analysis of observational data on the search for gravitational-lens candidates, shows a large number of pairs that could be explained by the complex geometry of the CS.
}

%%================================%%
%% Sample for structured abstract %%
%%================================%%




\maketitle

%%%%%%%%%%%%%%%%%%%%%%%%%%%%%%%%%%%%%%%%%%%%%%%
\section*{Introduction}\label{sec1}

Astronomers and physicists are closely approaching the search for nontrivial structures in the Universe, from topological defects to the consequences of the hidden multidimensionality of space-time. Such studies are actively supported by modern mathematical theory and existing gaps in understanding the unified picture of physical interactions and the structure of hidden sectors of matter: dark matter and dark energy.

Almost 50 years have passed since the prediction of cosmic strings (CS) as cosmic objects by T. W. B. Kibble \cite{3}-\cite{3_1}. CS were actively studied in subsequent works by Y. B. Zel'dovich \cite{4}, A. Vilenkin and others \cite{5}, \cite{6}, \cite{7}, \cite{8}, \cite{9}, \cite{10}, \cite{11}, \cite{12}, \cite{13}. In particular, the role of CS in the formation of gravitational-lens images was shown in \cite{5}, and the mechanism of generation of CMB anisotropy on CS was shown in \cite{14}.
The existence of CS does not contradict all currently available cosmological observational data and is widely supported in theory.  CS avoid the problems of the possible topological defects (single monopoles and domain walls), thus being the most interesting candidates from the point of view of observations. Variants of hybrid models (``dumbbells'', ``beads'', ``necklaces'' -- CS with monopoles at the ends and conglomerates of such structures) that do not contradict the observational data are also considered. They are particularly interesting, because, firstly, they are preferable from the point of view of theory \cite{27} -- \cite{85}, and secondly, they open up a wider area of space for their search. Indeed, a short dumbbell-type CS could be much closer to the observer than a CS of the same angular size, but ``piercing’’ the surface of the last scattering. Closer CS allow to search for more gravitational lensed galaxies. Hybrid models also appear in superstring theory \cite{28}.

The search for chains of gravitational lens events that a CS could form seems to be the most perspective astrophysical test. Firstly, it is possible to use data from numerous surveys and carry out such a search in automatic mode \cite{85}, and secondly, such a search complements the search for the CMB anisotropy. CS candidates, identified independently both in anisotropy data and by the presence of lens chains, are the most convincing.

The search for CS using gravitational lensed pair of distant objects began in the 1980s with the study of several pairs of quasars, \cite{34}. Numerous further unsuccessful attempts are described in the works \cite{35}, \cite{36},  \cite{37}, \cite{38}, \cite{39}, \cite{40}, \cite{41}, \cite{42}, \cite{43}, \cite{44}, \cite{45}, \cite{46}, \cite{47}, \cite{48}, \cite{49}, \cite{50}, \cite{51}. The search for observational manifestations of CS loops has been undertaken repeatedly, but also has not yet led to positive results.

Methods of searching for gravitational-lens images (forming chains, the so-called ``New Milky Way''), methods of searching for characteristic structures in the CMB anisotropy, as well as methods of gravitational-wave astronomy are applicable to a very wide class of CS predicted by the theory, being almost universal. 
But for simplicity of calculations, it is usually assumed that the CS is located perpendicular to the line of sight, then two images of distant source are formed in the plane of the CS lens, and are not rotated relative to each other.

The main goal of this paper is to demonstrate that in the case of a CS with a slope or a CS with a bend in the picture plane, the resulting images will be asymmetrical, with different positional angles. This fact significantly expands the search for gravitational lensed pairs.

The article is organized as follows. In the Chapter 1 we provide a brief introduction to gravitational lensing on a single straight CS. This based on recent work by \cite{72}. In the Chapter 2 we present the calculation of gravitation lensing effects due to a CS inclination. In the Chapter 3 we present the model of gravitational lensing on a curved CS. Thus in these two last chapters we consider the CS of general position with respect of the line of sight. We conclude by declaring the importance of consideration of the general position CS and discuss the new search strategy of gravitational lens pairs. After the Conclusion (Chapter 4) there are Appendixes. In the Appendix A we provide the flat approximation for a CS space-time with a conical singularity. In the Appendix B we provide the detailed calculations of energy-momentum tensor for a curved CS. In the Appendix C we describe photon trajectories for small metric perturbation. In the Appendix D we give derivation of a lens equation for a curved CS.

%%%%%%%%%%%%%%%%%%%%%%%%%%%%%%%%%%%%%%%%%%%%%%%
\section{Brief introduction to gravitational lensing on a single straight cosmic string}\label{sec2}

The metric of the cosmic string in cylindrical coordinates $(t, z, r, \varphi)$ has the well-known form:
    \begin{equation}\label{metric}
        g_{\mu\nu} = \text{diag}\Big(1, -1, -1, -r^2(1 - 4 G\mu)^2\Big) 
    \end{equation}

It is a conical metric with a deficit angle $\Delta\theta = 8\pi G\mu$ in a plane perpendicular to the string (see \ref{secA1}). The string lensing model in the flat geometric approximation is the following.

%(see fig. \ref{ris:image}).

  %  \begin{figure}[h!]
  %      \center{\includegraphics[width=0.8\linewidth]{s.png}}
  %      \caption{Lensing scheme of an object on a string in the simple geometric approximation (Fig. from \cite{48})}
  %      \label{ris:image}
  %  \end{figure}

    \begin{equation}\label{eq1}
        \begin{cases}
            \phi = -\eta + 4\pi G\mu \bigg(1 - \dfrac{R_s}{R_g}\bigg)\\
            \\
            \psi = \eta + 4\pi G\mu \bigg(1 - \dfrac{R_s}{R_g}\bigg)
        \end{cases}
    \end{equation}
where $R_g$ is the distance from an observer to a source (a galaxy), $R_s$ is the distance from an observer to the string,  $\eta$ is first coordinate, an angle between the direction to the string and the direction to the source ($\xi$ will be the second coordinate, the angle coordinate along the string).

\begin{figure}[h!]
    \centering
    \includegraphics[scale = 0.4]{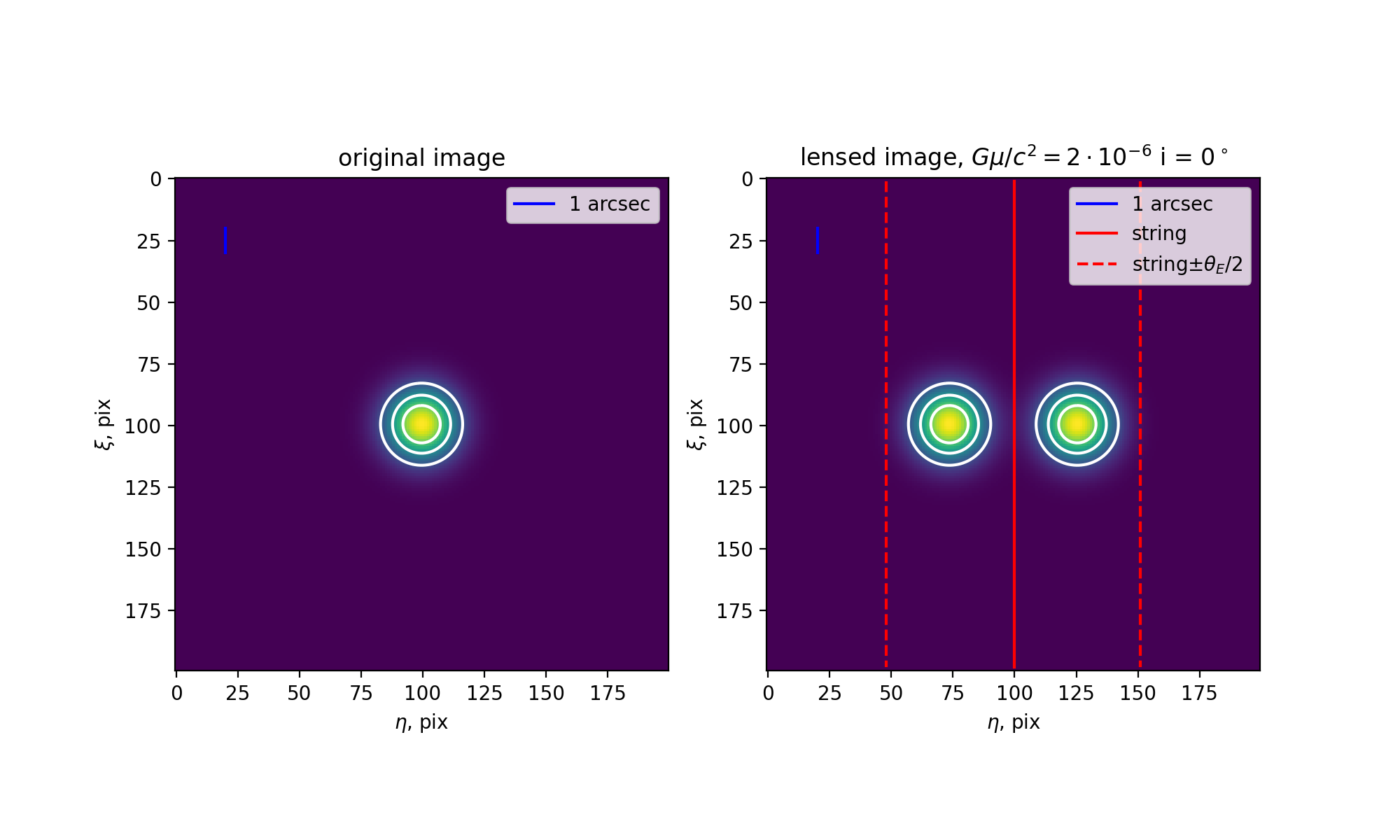}
    \caption{A simulation of a galaxy lensed by a cosmic string. The double image is clearly seen. Galaxy has been represented by a 2-dimensional gaussian distribution of intensity.}
    \label{fig:my_label1}
\end{figure}

If in (\ref{eq1}) $|\eta| < \theta_E$ we have two images of the source, where 
    \begin{equation}
        \theta_E = 8\pi G\mu \bigg(1 - \dfrac{R_s}{R_g}\bigg) \nonumber
    \end{equation}
The first image (shifted to the right of the string):
    \begin{equation}
        I_1(\eta, \xi) = 
        \begin{cases}
            I(\eta - \theta_E / 2, \xi),\: \eta > -\theta_E\\ \nonumber
            0,\: \eta \leq -\theta_E
        \end{cases}
    \end{equation}
The second image (shifted to the left of the string), (Fig. \ref{fig:my_label1}):
    \begin{equation}
        I_2(\eta, \xi) = 
        \begin{cases}
            I(\eta + \theta_E / 2, \xi),\: \eta < \theta_E\\ \nonumber
            0,\: \eta \geq \theta_E
        \end{cases}
    \end{equation}
or
    \begin{equation}
         I_{1+2}(\eta, \xi) = 
        \begin{cases}
            I(\eta + \theta_E / 2, \xi),\: \eta < -\theta_E\\
            I(\eta + \theta_E / 2, \xi) + I(\eta - \theta_E / 2, \xi),\: |\eta| \leq \theta_E \\
            I(\eta - \theta_E / 2, \xi),\: \eta > \theta_E \nonumber
        \end{cases}
    \end{equation}
what coincides with the results obtained earlier, \cite{48}.

%%%%%%%%%%%%%%%%%%%%%%%%%%%%%%%%%%%%%%%%%%%%%%%
\section{Effects due to a string inclination}\label{sec3}

Let us consider a case when cosmic string (CS) has an additional parameter, inclination $i > 0$ (an inclination of $0^\circ$ corresponds to an string located perpendicular to the line of sight, an inclination of $90^\circ$ corresponds to a string parallel to one). In this case the lensing parameter $\theta_E$  will be dependent on the position of the source, $\theta_E = \theta_E(i, \xi)$ due to two effects. 
    \begin{enumerate}
        \item for each $\xi$ the CS wil have different distance to an observer, $R_s = R_s(\xi)$,
        \item the effective deficit angle $\Delta \theta (i, \xi)$ for $i > 0$ is less than for $i = 0$.  
    \end{enumerate}

Let us discuss the first effect. For the triangle ``observer -- point on the string for $\xi = 0$ -- point on the string from the source'':
    \begin{equation}
        \frac{R_s(\xi)}{\sin(90^\circ - i)} = \frac{R_s(\xi = 0)}{\sin(180^\circ - \xi - (90^\circ - i))} \nonumber
    \end{equation}
    Then
    \begin{equation}
        R_s(\xi) = \frac{R_s(\xi = 0)}{\cos\xi + \tan i \sin \xi} \approx \frac{R_s(\xi = 0)}{1 + \xi \tan i } \nonumber
    \end{equation}
    
    \begin{equation}
        \theta_E = \Delta \theta (i) \bigg( 1 - \frac{R_s(\xi = 0)}{R_g (1 + \xi \tan i)}\bigg) \nonumber
    \end{equation}
    We can define the distance to the string $R_s$ to be $R_s(\xi = 0)$ to shorten the formulae.

    The deficit angle depends on the distance to the point of the string that we observe when passing the source. The conical metric is equivalent to the flat metric with cut, i.e. with two ``effective observers'' spaced by a distance of $L$:
    \begin{equation}\label{dtheta_condition}
        \Delta\theta \cdot h = L \nonumber
    \end{equation}
where $h$  is the length of the perpendicular from observer to the CS:
    \begin{equation}
        h = R_s \cos i \nonumber
    \end{equation}
    
\begin{figure}[h!]
        \centering
        \includegraphics[scale = 0.4]{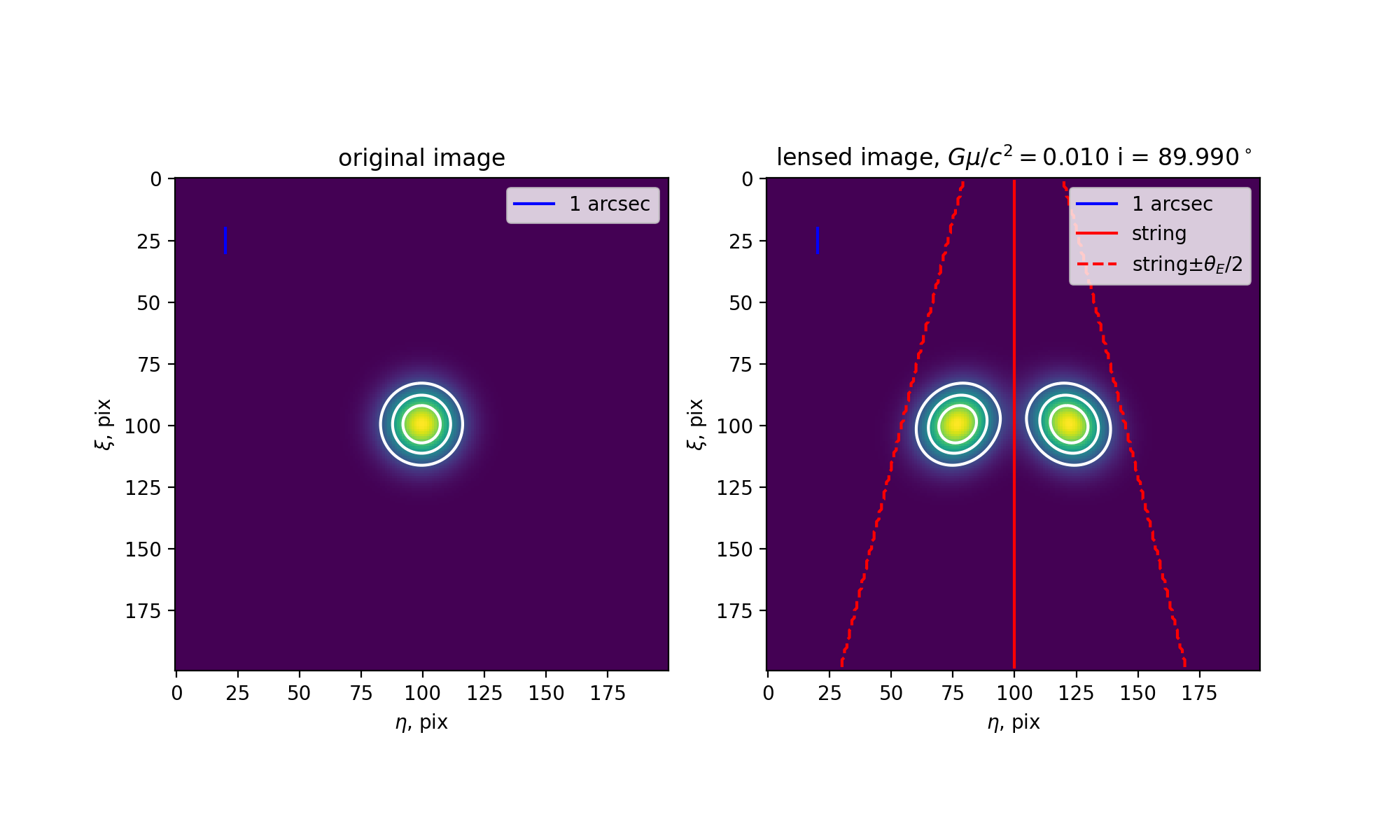}
        \caption{A simulation of a galaxy lensed by an inclined cosmic string. The difference between positional angles is clearly seen. Galaxy has been represented by a 2-dimensional gaussian distribution of intensity. The difference of positional angles of two images is visible.}
        \label{fig:my_label2}
    \end{figure}

\begin{figure}[h!]
        \centering
        \includegraphics[scale = 0.4]{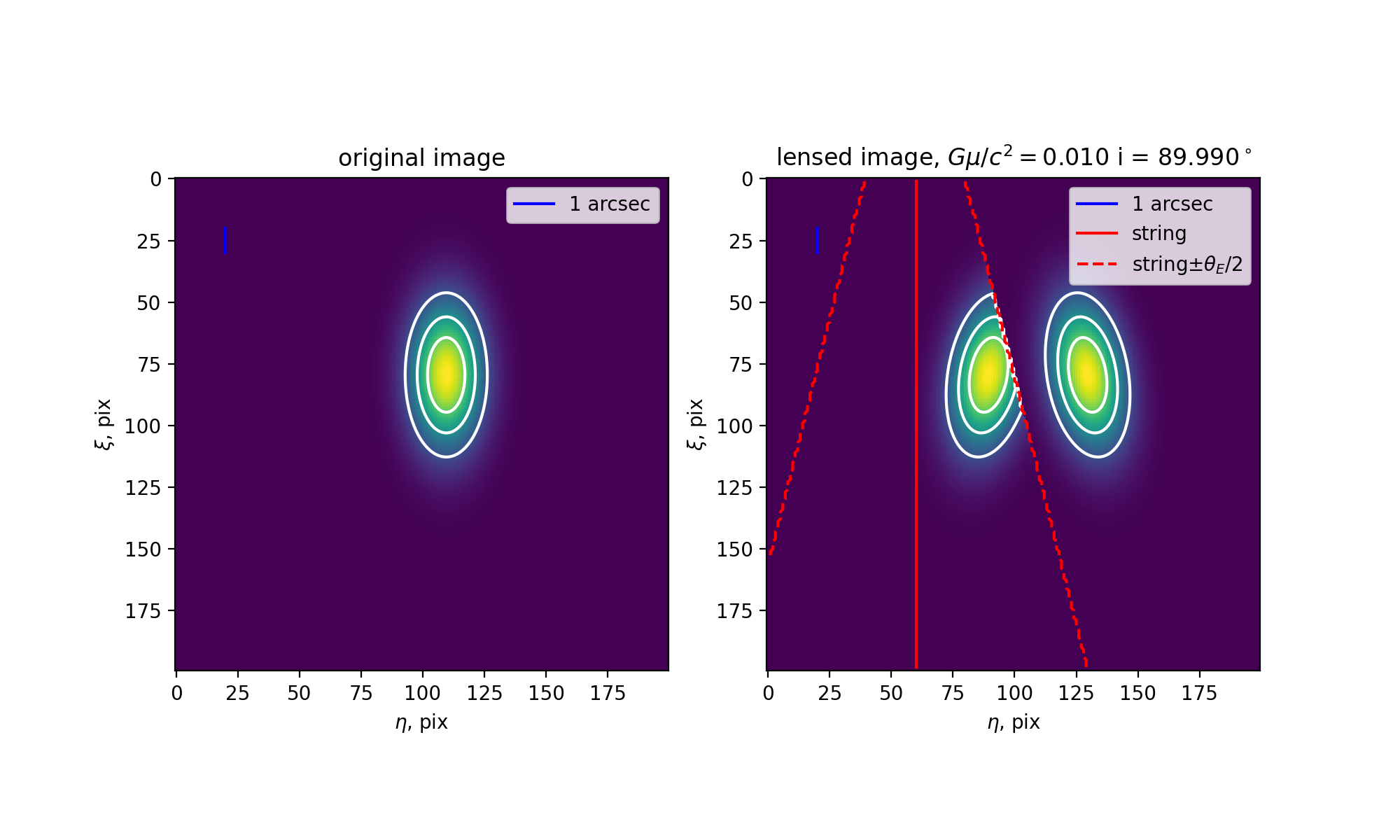}
        \caption{A simulation of a galaxy lensed by a cosmic string.  The difference between positional angles and an isophotes' cut are clearly seen. Galaxy has been represented by a 2-dimensional gaussian distribution of intensity. The difference of positional angles of two images is visible. The isophots' cuts are clearly visible.}
        \label{fig:my_label3}
    \end{figure}

    In case of inclined string we can adopt the same geometric framework, but in 3 dimensions. For every $\xi$ the picture will be the same except now $\Delta\theta(i, \xi)$ is defined by equation (\ref{dtheta_condition}). Also in this equation instead of $h$ we have $R_s(\xi)$. Since $L$ is the same for all such $\xi$, then
    \begin{equation}
        \Delta\theta R_s \cos i = \Delta \theta (i, \xi) R_s(\xi) \nonumber
    \end{equation}
    and
    \begin{equation}
        \Delta \theta (i, \xi) = \Delta\theta\: (\cos i + \xi \sin i) \nonumber
    \end{equation}
    Finally,
    \begin{equation}
        \theta_E(i, \xi) = \Delta\theta\: (\cos i + \xi \sin i)) \bigg( 1 - \frac{R_s}{R_g (1 + \xi \tan i)}\bigg) \nonumber
    \end{equation}
    All the other steps in constructing a lensing transformation are the same, so we end up with (see Fig. \ref{fig:my_label2}, \ref{fig:my_label3}):
    \begin{equation}
        I_{1+2}(\eta, \xi) = 
        \begin{cases}
            I(\eta + \theta_E(i, \xi) / 2, \xi),\: \eta < -\theta_E(i, \xi)\\
            I(\eta + \theta_E(i, \xi) / 2, \xi) + I(\eta - \theta_E(i, \xi) / 2, \xi),\: |\eta| \leq \theta_E(i, \xi) \\
            I(\eta - \theta_E(i, \xi) / 2, \xi),\: \eta > \theta_E(i, \xi) \nonumber
        \end{cases}
    \end{equation}

    Since $\xi \ll 1$, it is more useful to make an expansion of $\theta_E(i, \xi)$:
    \begin{equation}
        \theta_E(i, \xi) = \theta_E(i, \xi = 0) + \frac{\partial \theta_E}{\partial \xi}\bigg|_{\xi = 0} \cdot \xi \nonumber
    \end{equation}
    where
    \begin{equation}
        \theta_E(i, \xi = 0) = \Delta\theta\: \cos i \bigg( 1 - \frac{R_s}{R_g} \bigg) \nonumber
    \end{equation}
    \begin{equation}
        \frac{\partial \theta_E}{\partial \xi}\bigg|_{\xi = 0} = \Delta \theta \sin i \nonumber
    \end{equation}

   Note that from here the rate of increase of $\theta_E$ is limited by the value of $\Delta\theta$ in radians. Knowing the limitations on it ($\Delta \theta \lesssim 10^{-5}$), the tilt limitations during lensing can be neglected in almost all realistic cases.

   It should also be borne in mind that such a picture may occur, even if only a part of the string on the line of sight will have a large slope.

%%%%%%%%%%%%%%%%%%%%%%%%%%%%%%%%%%%%%%%%%%%%%%%
\section{Model of gravitational lensing on a curved string}\label{sec4}

    In the previous paragraph we have discussed the effects of inclination on a picture one can get from an object that lies behind the string. Now the model can be developed further.

    If a CS on a line of sight has a bend with an angle $\theta \neq 0^\circ$, the metric does not have the form (\ref{metric}). To get the picture of a galaxy behind the CS one should calculate the geodesics, using the general relativity framework. 

    Consider a static CS, that consists of 2 straight lines, connected by a Besier curve (see Fig. \ref{ris:image}). As a matter of simplicity, the string is located in the $(x^1, x^2)$ plane, since the effect of inclination (see Chapter \ref{sec3}) can be neglected:

    \begin{equation}
    \begin{cases}
            X^1(s) = x(s) = 
            \begin{cases}
                0, \;s \in (-\infty, - R)\\
                \dfrac{R \sin \theta}{4} \bigg(1 + \dfrac{s}{R} \bigg)^2, \;s \in [-R, R]\\
                s \sin \theta, \;s \in (R, +\infty)
            \end{cases}
            \\
            X^2(s) = y(s) = 
            \begin{cases}
                s, \;s \in (-\infty, - R)\\
                \dfrac{R \cos \theta}{4} \bigg(1 + \dfrac{s}{R} \bigg)^2 -\dfrac{R}{4} \bigg(1 - \dfrac{s}{R} \bigg)^2, \;s \in [-R, R]\\
                s \cos \theta, \;s \in (R, +\infty)
            \end{cases}
            \\
            X^3(s) = z(s) = R_g - R_s \nonumber
    \end{cases}
    \end{equation}

    The angle $\theta$ describes how much the string changed the positional angle and $R$ is the characteristic length of such bend. After we have defined the string location, we need to compute the energy-momentum tensor (EMT) of such string, using the standard formula:

    \begin{equation}
        T_{\mu\nu}(\mathbf{x}) = \mathcal{F}^{-1}\bigg\{\mu \int_{-\infty}^{+\infty} d\sigma \exp(- i \mathbf{kX}) \Big( \partial_{t}X_\mu \partial_{t}X_\nu - \partial_{\sigma}X_\mu\partial_{\sigma}X_\nu\Big)\bigg\} \nonumber
    \end{equation}
    where the $\mathcal{F}^{-1}$ is the inverse Fourier transformation frequency domain $\mathbf{k}$ to coordinate domain $\mathbf{x}$. The complete derivation of EMT can be seen in the Appendix \ref{secA2}. If we assume that the bend is small, compared to other parts of the string ($R = 0$), EMT can be written as:
    \begin{equation}
    T_{\mu\nu} = 
        \mu \left(
        \begin{array}{cccc}
            \delta_{\downarrow} + \delta_{\uparrow} & 0 & 0 & 0 \\
            0 & - \delta_{\uparrow}\sin^2\theta & - \delta_{\uparrow}\sin\theta\cos\theta  & 0 \\
            0 & - \delta_{\uparrow}\sin\theta \cos\theta & -\delta_{\downarrow} - \delta_{\uparrow} \cos^2\theta  & 0\\
            0 & 0 & 0 & 0 \\
        \end{array}
        \right) \nonumber
    \end{equation}
    where:
    \begin{equation}
        \delta_{\downarrow} = \delta(z)\delta(x)\Big(1 - H(y) \Big) \nonumber
    \end{equation}
    \begin{equation}
        \delta_{\uparrow} = \delta(z)\delta(x\cos\theta - y\sin\theta)H(x\sin\theta + y\cos\theta) \nonumber
    \end{equation}
    and $H(x)$ is a unit step Heaviside function. 
    
    According to the linearized Einstein equation, nonzero components of the EMT give us the nonzero components of the metric perturbation $h_{\mu\nu}$. The metric itself is divergent if we assume the infinitely thin CS, but the Cristoffel symbols and the geodesics equation for photons in such space can be derived. The full derivation can be found in Appendixes \ref{secA3} and \ref{secA4}. The final result of this section is an initial boundary problem (IBP) for photon trajectory, that should be solved numerically in order to get the lensed picture:

    \begin{equation*}
    \begin{cases}
        \dfrac{d \mathbf{v}}{da} =  -\dfrac{1}{2}
        \nabla_\mathbf{n}
        (h_\uparrow + h_\downarrow) -  
        \bigg[\dfrac{\partial h_\uparrow}{\partial a}
        \left(
        \begin{array}{cc}
            \cos^2\theta & \sin\theta\cos\theta \\
            \sin\theta\cos\theta & \sin^2\theta
        \end{array}
        \right)
        + \dfrac{\partial h_\downarrow}{\partial a}
        \left(
        \begin{array}{cc}
            1 & 0 \\
            0 & 0
        \end{array}
        \right)\bigg]
        \mathbf{v}\\
        d\mathbf{n}/da = \mathbf{v}
        \\
        \mathbf{n}(a = 0) = \mathbf{n}_0
        \\
        \mathbf{n}(a = 1) =0
    \end{cases}
    \end{equation*}
    where $\mathbf{n}_0$ is the initial direction to the point source. The definitions for $a, \mathbf{n}, h_{\uparrow, \downarrow}$ and the partial derivatives of functions $h_{\uparrow, \downarrow}$ can be seen in the Appendix \ref{secA4}.
    
\begin{figure}[h!]
        \centering
        \includegraphics[scale = 0.5]{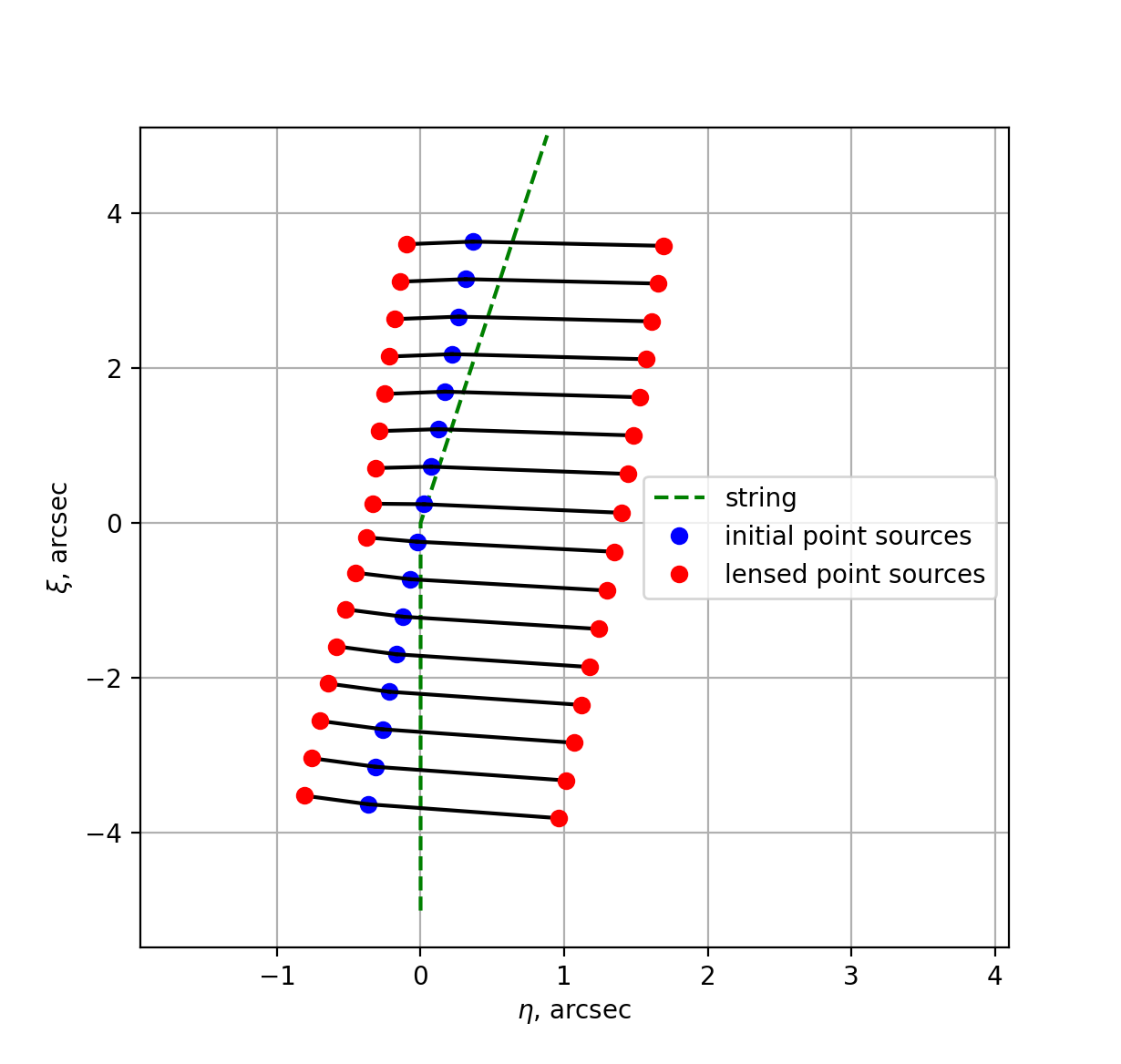}
        \caption{An example of lensing of 16 point sources (blue dots) on a string with a bend $\theta = 10^\circ$, tension $G\mu = 7.0 \cdot 10^{-7}$ and $R_s/R_g = 0.5$. The new positions of the sources are marked by red dots, the string position is drawn by a green dashed line.}
        \label{fig:curved10}
    \end{figure}

\begin{figure}[h!]
        \centering
        \includegraphics[scale = 0.5]{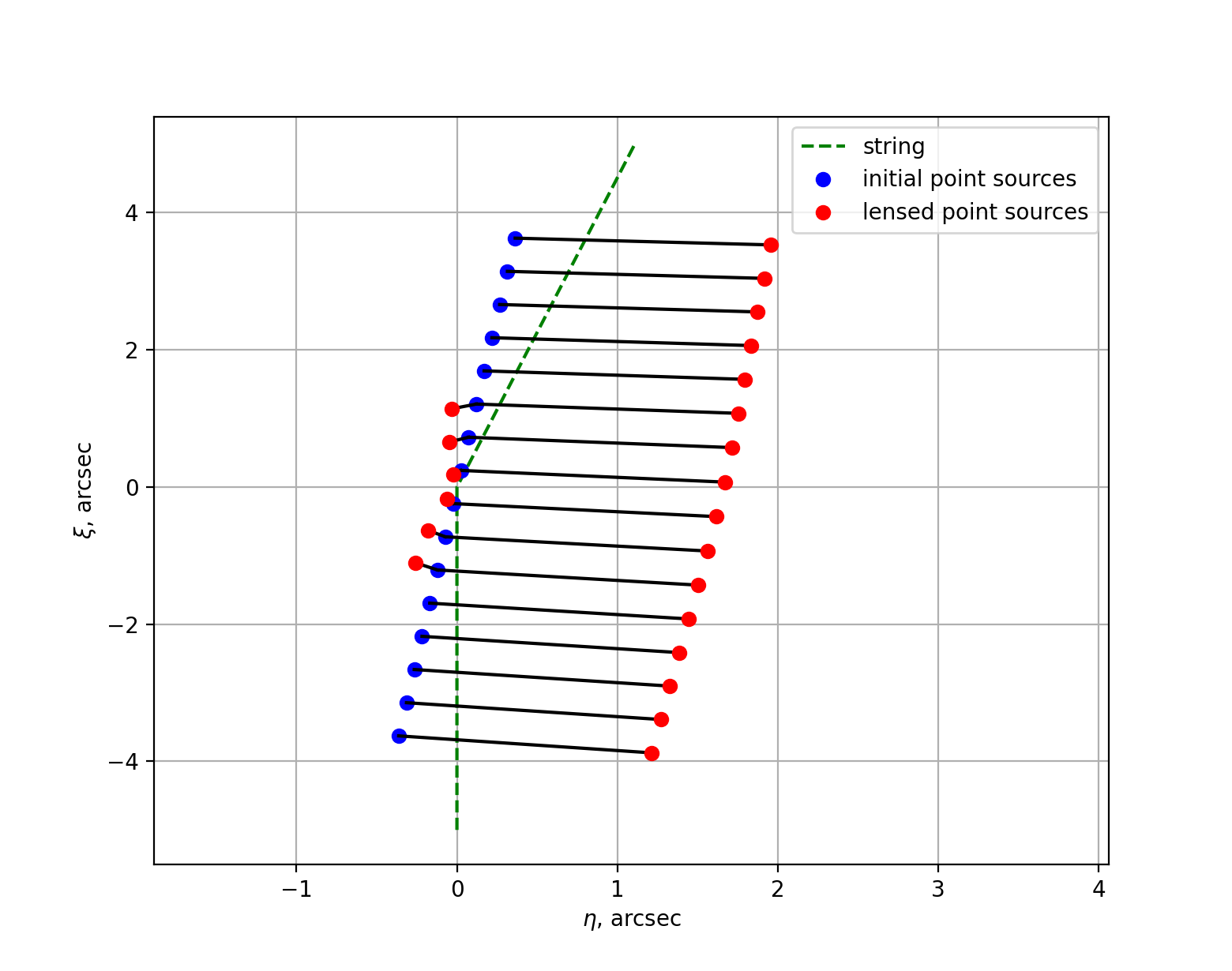}
        \caption{An example of lensing of 16 point sources (blue dots) on a string with a bend $12.5^{\circ}$, tension $G\mu = 7.0 \cdot 10^{-7}$ and $R_s/R_g = 0.5$. The new positions of the sources are marked by red dots, the string position is drawn by a green dashed line.}
        \label{fig:curved11}
    \end{figure}

    After getting the solution of IBP we can calculate $\mathbf{v}_f = -\mathbf{v}(a = 1)$ - the direction under which the light from the point source $\mathbf{n}_0$ is seen. Since the IBP can have several solutions, we can numerically check the area around $\mathbf{n}_0$ using the shooting method to formally obtain a map $-\mathbf{v}_f(\mathbf{n}_0)$. After that the lensing picture can also be computed numerically. The result of this computation can be seen in the Fig. (\ref{fig:curved10}), (\ref{fig:curved11}).

\begin{figure}[h!]
        \centering
        \includegraphics[scale = 0.5]{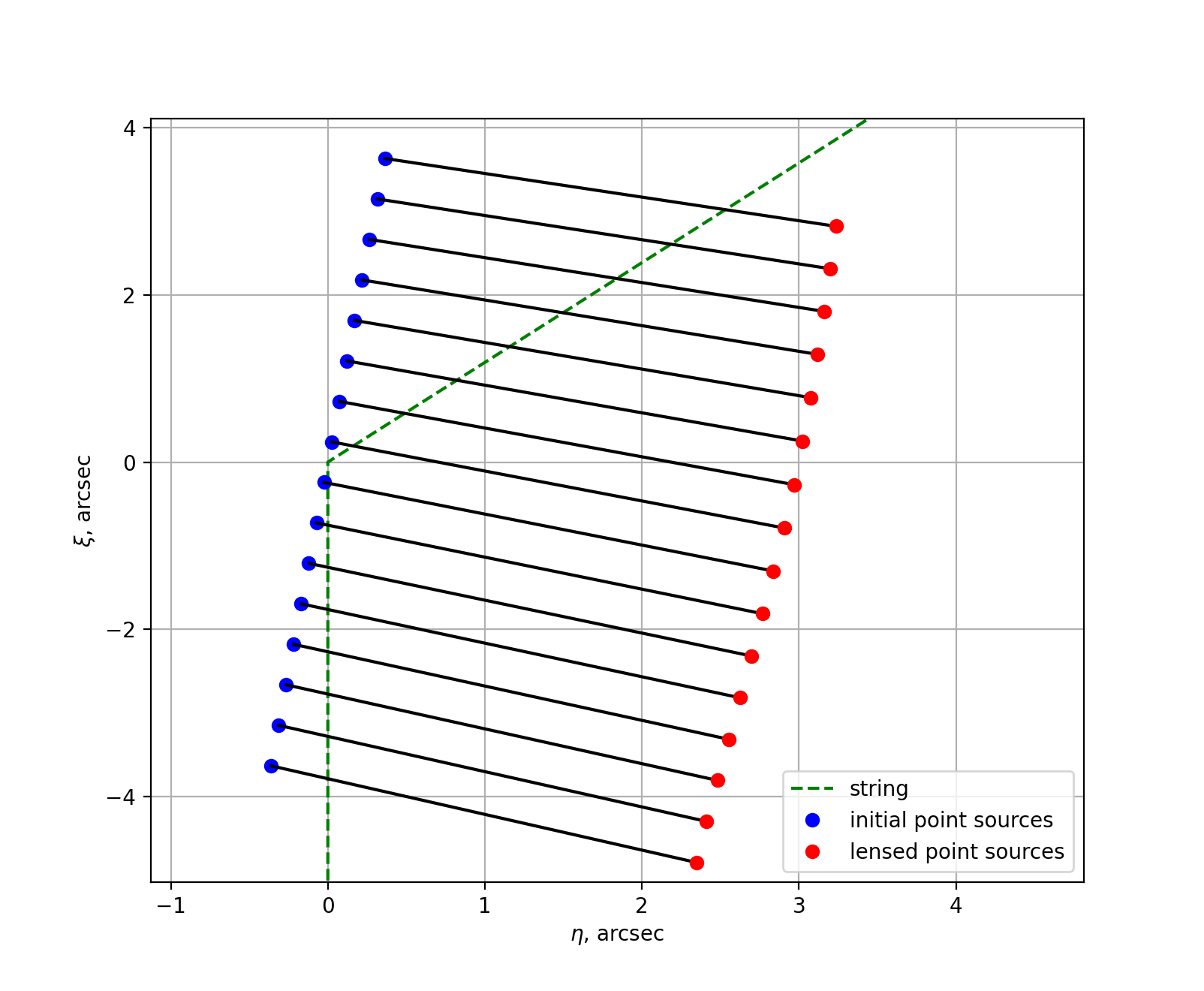}
        \caption{An example of lensing of 16 point sources (blue dots) on a string with a bend $\theta = 40^\circ$, tension $G\mu = 7.0 \cdot 10^{-7}$ and $R_s/R_g = 0.5$. The new positions of the sources are marked by red dots, the string position is drawn by a green dashed line.}
        \label{fig:curved40}
    \end{figure}

    One can also notice, that for big angle of a bend the double image cannot be obtained even for an object that lies behind the string (see Fig. \ref{fig:curved40}). From numerical simulations the critical angle is $\sim 13^{\circ}$. This result can be an argument for the lack of double images of galaxies.

    It was also shown in \cite{deLaix:1997jt} and \cite{deLaix:1997dj} that CS can produce more than 2 images of distant sources, due to their small scale stricture, which can be the case in our model, if  one incorporates more than one bend. However, the modern limitations on the deficit angle are of same magnitude, as our angular resolution in visible light. This means that GL events on CS with more than 2 images are difficult to find and analyze.

%%%%%%%%%%%%%%%%%%%%%%%%%%%%%%%%%%%%%
%\section{Discussion}\label{sec5}

%%%%%%%%%%%%%%%%%%%%%%%%%%%%%%%%%%%%%
\section{Conclusions}\label{sec6}

The model presented in the paper first generalizes the gravitational lensing on a CS of general position. Gravitational lensing on a CS with an inclination in the plane coinciding with the beam of vision is considered. Gravitational lensing on a CS curved in a plane perpendicular to the line of sight is considered. In the work, the curvature of the CS is given by one fracture with a given angle. A fundamentally new result is that the bending of the CS crucially affects the number of images. So, with a larger value of the bending angle (approximately more than $13^{\circ}$), the second image disappears, which can serve as an argument to explain the absence of a large number of  gravitational-lens chains (new ''Milky Ways``).

The simulation results are applied by authors to the analysis of gravitational-lens candidates in the field of the previously found by CMB analysis \cite{sazhina} candidate string CSc-1 (in preparation).

For cosmology, astrophysics, and theoretical physics the discovery of CS is without a doubt a huge step in understanding the structure of the Universe, especially its global properties and the earliest stages of its evolution. 

The presented in this paper detailed theoretical study opens up fundamentally new ways to search for gravitational-lens events, which, together with the analysis of CMB anisotropy, will allow statistically significant detection of CS.

%%%%%%%%%%%%%%%%%%%%%%%%%%%%%%%%%%%%%
\bmhead{Data availability}

The code for the modeling of lensing by the inclined and bended string can be sent upon request by e-mail.

%%%%%%%%%%%%%%%%%%%%%%%%%%%%%%%%%%%%%
\bmhead{Acknowledgments}

I.I. Bulygin acknowledges the financial support by the <<BASIS>> foundation and expresses its gratitude to the <<Traektoria>> foundation.

%%%%%%%%%%%%%%%%%%%%%%%%%%%%%%%%%%%%%
\begin{appendices}

\section{Flat approximation for a CS space-time with a conical singularity }\label{secA1}

Geodesic trajectory has the form:
    \begin{equation}
        \frac{d^2 x^\lambda}{ds^2} = - \Gamma^{\lambda}_{\mu\nu} \frac{d x^\mu}{ds} \frac{d x^\nu}{ds} \nonumber
    \end{equation}
where
    \begin{equation}
        \Gamma^{\lambda}_{\mu\nu}=\frac{g^{\lambda \alpha}}{2}\bigg(\frac{\partial g_{\mu \alpha}}{\partial x^\nu} + \frac{\partial g_{\nu \alpha}}{\partial x^\mu} - \frac{\partial g_{\mu\nu}}{\partial x^\alpha}\bigg) \nonumber
    \end{equation}

Nonzero derivatives in the metric of the cosmic string only has $g_{\varphi\varphi}$ component, so only non-zero Christoffel symbols are:
    \begin{equation}
        \Gamma^{\varphi}_{\varphi r} = \frac{g^{\varphi \varphi}}{2} \frac{\partial g_{\varphi \varphi}}{\partial r} = \frac{1}{r} \nonumber
    \end{equation}

    \begin{equation}
        \Gamma^{\varphi}_{r \varphi} = \frac{g^{\varphi \varphi}}{2} \frac{\partial g_{\varphi \varphi}}{\partial r} = \frac{1}{r} \nonumber
    \end{equation}

    \begin{equation}
        \Gamma^{r}_{\varphi \varphi} = r (1 - 4 G \mu)^2 \nonumber
    \end{equation}

The first two characters are the same for a flat space and do not contain string parameters. This means that we can consider the space to be flat. The third character is used in combination with the derivatives $d\varphi/ds$. If we make a replacement:

    \begin{equation}
        \varphi' = \varphi (1 - 4 G \mu) \nonumber
    \end{equation}

Then all the equations will take the form as for a flat conical space with deficit angle $\Delta \theta = 8\pi G \mu$ in the plane perpendicular to the string. 

\section{Energy-momentum tensor for a curved string}\label{secA2}

To define the EMT of the curved string, we assume the area on which the galaxy is lensed to be small. So only one segment with nonzero curvature is needed for the model. Thus we approximate the string with two straigth lines connected by Besier curve:
\begin{equation}
    \begin{cases}
            X^1(s) = x(s) = 
            \begin{cases}
                0, \;s \in (-\infty, - R)\\
                \dfrac{R \sin \theta}{4} \bigg(1 + \dfrac{s}{R} \bigg)^2, \;s \in [-R, R]\\
                s \sin \theta, \;s \in (R, +\infty)
            \end{cases}
            \\
            X^2(s) = y(s) = 
            \begin{cases}
                s, \;s \in (-\infty, - R)\\
                \dfrac{R \cos \theta}{4} \bigg(1 + \dfrac{s}{R} \bigg)^2 -\dfrac{R}{4} \bigg(1 - \dfrac{s}{R} \bigg)^2, \;s \in [-R, R]\\
                s \cos \theta, \;s \in (R, +\infty)
            \end{cases}
            \\
            X^3(s) = z(s) = R_g - R_s\\
    \end{cases} \nonumber
\end{equation}
In this article we use the convention $X^0 = t$ and assume the static string.

\begin{figure}[h!]
        \center{\includegraphics[width=0.5\linewidth]{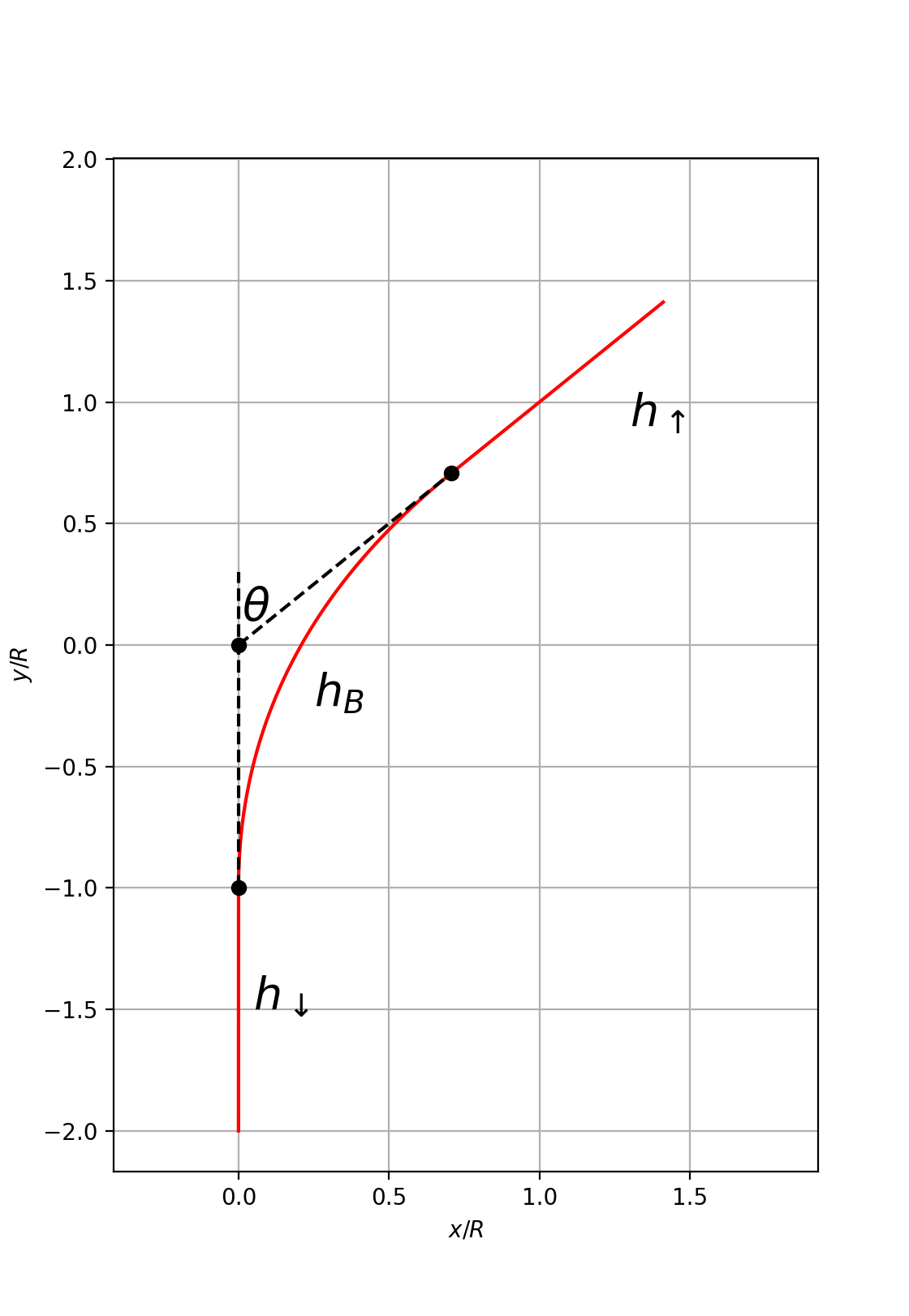}}
        \caption{String position in $(x^1, x^2)$ plane}
        \label{ris:image}
\end{figure}

The parameter $s$ in string location $\mathbf{X}$ is not its natural parameterization $\sigma$, which is used in calculation of EMT:
\begin{equation}
     T_{\mu\nu}(\mathbf{x}) = \mathcal{F}^{-1}\bigg\{\mu \int_{-\infty}^{+\infty} d\sigma \exp(- i \mathbf{kX}) \Big( \partial_{t}X_\mu \partial_{t}X_\nu - \partial_{\sigma}X_\mu\partial_{\sigma}X_\nu\Big)\bigg\} \nonumber
\end{equation}
The connection between $s$ and $\sigma$ is:
\begin{equation}
     \frac{d\sigma}{ds} = \rho(s) = \sqrt{x'(s)^2 + y'(s)^2 + z'(s)^2 }  = 
        \begin{cases}
            1,\;s \in (-\infty, - R)\\
            \sqrt{\cos^2 \theta/2 + (s/R)^2 \sin^2\theta / 2}, \;s \in [-R, R]\\
            1, \;s \in (R, +\infty)\\
        \end{cases} \nonumber
\end{equation}
This connection transforms the EMT into:
\begin{equation}
    T^{\mu\nu}(\mathbf{x}) =  \mu \int_{-\infty}^{+\infty} \rho(s) ds\:  \Big( \partial_{t}X^\mu \partial_{t}X^\nu - \partial_{\sigma}X^\mu\partial_{\sigma}X^\nu\Big) \delta^{(3)}\Big(\mathbf{x} - \mathbf{X}(s)\Big) \nonumber
\end{equation}
The main step in calculation of EMT is to split the integral into 3 parts: 
\begin{itemize}
    \item $\downarrow$ - for $s \in (-\infty, - R)$, where the positional angle is 0;
    \item $\uparrow$ - for $s \in (R, +\infty)$, where the positional angle is $\theta$;
    \item $B$ - for $s \in [-R, R]$, where the bend is located.
\end{itemize}
and then find integrals that describe all components of the EMT in some linear combination, for example:
\begin{equation}
    T^{00}(\mathbf{x}) =  \mu \int_{-\infty}^{-R} ds\:\delta^{(3)}\Big(\mathbf{x} - \mathbf{X}(s)\Big) + \nonumber
\end{equation}
\[
+ \mu \int_{-R}^{R} ds\:\sqrt{\cos^2 \theta/2 + (s/R)^2 \sin^2\theta / 2} \times \delta^{(3)}\Big(\mathbf{x} - \mathbf{X}(s)\Big) +
\]
\[
 + \mu \int_{R}^{+\infty} ds\: \delta^{(3)}\Big(\mathbf{x} - \mathbf{X}(s)\Big) = 
\]
\[
= \mu (\delta_{\downarrow} + \Delta_{00}^{B} + \delta_{\uparrow})
\]
where the calculated integrals are respectfully:
\begin{equation}
    \delta_{\downarrow} = \delta(z)\delta(x)\Big(1 - H(y + R) \Big) \nonumber
\end{equation}
\begin{equation}
    \Delta_{00}^{B} = \int_{-R}^{R} ds\:\sqrt{\cos^2 \theta/2 + (s/R)^2 \sin^2\theta / 2} \times \delta^{(3)}\Big(\mathbf{x} - \mathbf{X}(s)\Big) \nonumber
\end{equation}
\begin{equation}
    \delta_{\uparrow} = \delta(z)\delta(x\cos\theta - y\sin\theta)H(x\sin\theta + y\cos\theta - R) \nonumber
\end{equation}
Some components are trivially zero, such as:
\begin{equation}
    T^{0i} = T^{i0} = T^{3\mu} = T^{\mu 3} = 0 \nonumber
\end{equation}
The other ones include $\partial_\sigma = \rho(s)^{-1} \partial_s$:
\begin{equation}
    T^{11} (\mathbf{x}) =  0 - \nonumber
\end{equation}
\[
 - \frac{\mu}{4} \int_{-R}^{R} ds\:\frac{\sin^2\theta \bigg( 1 + \dfrac{s}{R}\bigg)^2}{\sqrt{\cos^2 \theta/2 + (s/R)^2 \sin^2\theta / 2}} \times \delta^{(3)}\Big(\mathbf{x} - \mathbf{X}(s)\Big) - 
\]
\[
  - \mu \sin^2\theta \int_{R}^{+\infty} ds\: \delta^{(3)}\Big(\mathbf{x} - \mathbf{X}(s)\Big) = 
\]
\[
= - \mu ( \Delta_{11}^{B} + \delta_{\uparrow}\sin^2\theta )
\]

\begin{equation}
    T^{12} (\mathbf{x}) =  0 - \nonumber
\end{equation}
\[
 - \frac{\mu}{4}\int_{-R}^{R} ds\:\frac{\bigg[ \bigg(1 + \dfrac{s}{R}\bigg) \cos \theta + \bigg(1 - \dfrac{s}{R}\bigg)\bigg]\bigg(1 + \dfrac{s}{R}\bigg) \sin \theta }{\sqrt{\cos^2 \theta/2 + (s/R)^2 \sin^2\theta / 2}} \times \delta^{(3)}\Big(\mathbf{x} - \mathbf{X}(s)\Big) - 
\]
\[
  - \mu \cos\theta\sin\theta \int_{R}^{+\infty} ds\: \delta^{(3)}\Big(\mathbf{x} - \mathbf{X}(s)\Big)
\]
\[
= - \mu ( \Delta_{12}^{B} + \delta_{\uparrow}\sin\theta\cos\theta )
\]

\begin{equation}
    T^{22} (\mathbf{x}) =  -\mu\int_{-\infty}^{-R} ds\:\delta^{(3)}\Big(\mathbf{x} - \mathbf{X}(s)\Big) - \nonumber
\end{equation}
\[
 - \frac{\mu}{4} \int_{-R}^{R} ds\:\frac{\bigg[ \bigg(1 + \dfrac{s}{R}\bigg) \cos \theta + \bigg(1 - \dfrac{s}{R}\bigg)\bigg]^2}{\sqrt{\cos^2 \theta/2 + (s/R)^2 \sin^2\theta / 2}} \times \delta^{(3)}\Big(\mathbf{x} - \mathbf{X}(s)\Big) - 
\]
\[
  - \mu \cos^2\theta\int_{R}^{+\infty} ds\: \delta^{(3)}\Big(\mathbf{x} - \mathbf{X}(s)\Big)
\]
\[
= -\mu (\delta_{\downarrow} + \Delta_{22}^{B} + \delta_{\uparrow} \cos^2\theta )
\]

All the integrals named $\Delta_{ij}^{B}$ are of the form:
\begin{equation}
    \delta(z) \int_{-R}^{R} f(s) \delta\big(x_0 - x(s)\big) \delta\big(y_0 - y(s)\big) ds = 
    \begin{cases}
        \dfrac{f(s_0)\delta(z)\delta\big(y_0 - y(s_0)\big)}{|x'(s_0)|}\bigg|_{x_0 = x(s_0)}, \:\text{if}\: y_0 = y(x_0)\\
        0, \:\text{if}\: y_0 \neq y(x_0)
    \end{cases} \nonumber
\end{equation}
Thus the exact solutions for $\Delta_{ij}^{B}$ are:
% добавить плотности и ступеньки хевисайда! А 
\begin{equation}
    \Delta_{00}^{B} = \delta(z) \delta\bigg(y + x \tan\frac{\theta}{2} - 2\sqrt{\frac{xR}{\sin\theta}} + R \bigg) \sqrt{\dfrac{\cos^2\dfrac{\theta}{2} + \sin^2\dfrac{\theta}{2} \bigg(2\sqrt{\dfrac{x}{R\sin\theta}} - 1\bigg)^2}{\dfrac{x\sin\theta}{R}}} \times \nonumber
\end{equation}
\[
\times (1 - H(x\sin\theta + y\cos\theta - R)) H(y + R)
\]
\begin{equation}
    \Delta_{11}^{B} = \delta(z) \delta\bigg(y + x \tan\frac{\theta}{2} - 2\sqrt{\frac{xR}{\sin\theta}} + R \bigg) \sqrt{\dfrac{\dfrac{x\sin\theta}{R}}{\cos^2\dfrac{\theta}{2} + \sin^2\dfrac{\theta}{2} \bigg(2\sqrt{\dfrac{x}{R\sin\theta}} - 1\bigg)^2}} \times \nonumber
\end{equation}
\[
\times (1 - H(x\sin\theta + y\cos\theta - R)) H(y + R)
\]
\begin{equation}
    \Delta_{12}^{B} = \delta(z) \delta\bigg(y + x \tan\frac{\theta}{2} - 2\sqrt{\frac{xR}{\sin\theta}} + R \bigg) \dfrac{1 - \sqrt{\dfrac{x}{R\sin\theta}} (1 - \cos\theta)}{\sqrt{\cos^2\dfrac{\theta}{2} + \sin^2\dfrac{\theta}{2} \bigg(2\sqrt{\dfrac{x}{R\sin\theta}} - 1\bigg)^2}} \times \nonumber
\end{equation}
\[
\times (1 - H(x\sin\theta + y\cos\theta - R)) H(y + R)
\]
\begin{equation}
    \Delta_{22}^{B} = \delta(z) \delta\bigg(y + x \tan\frac{\theta}{2} - 2\sqrt{\frac{xR}{\sin\theta}} + R \bigg) \dfrac{\sqrt{\dfrac{R}{x \sin\theta}}\bigg(1 - \sqrt{\dfrac{x}{R\sin\theta}} (1 - \cos\theta)\bigg)^2}{\sqrt{\cos^2\dfrac{\theta}{2} + \sin^2\dfrac{\theta}{2} \bigg(2\sqrt{\dfrac{x}{R\sin\theta}} - 1\bigg)^2}} \times \nonumber
\end{equation}
\[
\times (1 - H(x\sin\theta + y\cos\theta - R)) H(y + R)
\]
So the full answer is:
\begin{equation}
    T_{\mu\nu} = 
        \mu \left(
        \begin{array}{cccc}
            \delta_{\downarrow} + \Delta_{00}^{B} + \delta_{\uparrow} & 0 & 0 & 0 \\
            0 & -\Delta_{11}^{B} - \delta_{\uparrow}\sin^2\theta & -\Delta_{12}^{B} - \delta_{\uparrow}\sin\theta\cos\theta  & 0 \\
            0 & -\Delta_{12}^{B} - \delta_{\uparrow}\sin\theta\cos\theta & -\delta_{\downarrow} - \Delta_{22}^{B} - \delta_{\uparrow} \cos^2\theta  & 0\\
            0 & 0 & 0 & 0 \\
        \end{array}
        \right) \nonumber
\end{equation}
To solve the linearized Einstein's equations we also need the source function. In the approximation of small bend ($R \ll R_s \Delta \theta $ or simply $R = 0$):
\begin{equation}
    S_{\mu\nu} = 
        \mu \left(
        \begin{array}{cccc}
            0 & 0 & 0 & 0 \\
            0 & -\delta_{\downarrow} -\delta_{\uparrow}\cos^2\theta & - \delta_{\uparrow}\sin\theta\cos\theta  & 0 \\
            0 & - \delta_{\uparrow}\sin\theta\cos\theta &  -\delta_{\uparrow} \sin^2\theta  & 0\\
            0 & 0 & 0 & \delta_{\downarrow} + \delta_{\uparrow} \\
        \end{array}
        \right) \nonumber
\end{equation}

\section{Photon trajectories for small metric perturbation}\label{secA3}

Let $x^1$, $x^2$ be the axis parallel to the picture plane, $x^3$ be the axis parallel to the line of sight. The source (galaxy) will be at $x^3 = 0$, the observer will be located at $x^3 = R_g$. The metric perturbation (in the article it is the curved cosmic string) between the observer and the source will be placed at $x^3 = R_g - R_s$ and it will be in the form:

    \begin{equation}
        h_{\mu\nu} = 
        \left(
        \begin{array}{cccc}
            0 & 0 & 0 & 0 \\
            0 & h_{11} & h_{12} & 0 \\
            0 & h_{12} & h_{22} & 0\\
            0 & 0 & 0 & h_{33} \\
        \end{array}
        \right) \nonumber
    \end{equation}

Since photon travel along the null geodesics, it is convenient to choose time $x^0 = t$ as a parameter:
    \begin{equation}
        \frac{d^2 x^i}{dt^2} = - \bigg( \Gamma^i_{\mu\nu} - \Gamma^0_{\mu\nu} \frac{d x^{i}}{dt} \bigg) \frac{d x^{\mu}}{dt} \frac{d x^{\nu}}{dt}, \;i = 1, 2, 3 \nonumber
    \end{equation}
In the weak field approximation one can write:
\begin{equation}
    \Gamma^{\lambda}_{\mu\nu} \approx \frac{1}{2}\bigg(\frac{\partial h_{\mu}^{.\:\lambda}}{\partial x^\nu} + \frac{\partial h_{\nu}^{.\:\lambda}}{\partial x^\mu} - \frac{\partial h_{\mu\nu}}{\partial x_\lambda}\bigg) \nonumber
\end{equation}
It is easy to see that for this metric perturbation:  
    \begin{equation}
        \Gamma^0_{\mu\nu} = \frac{1}{2}\bigg(\frac{\partial h_{\mu}^{.\:0}}{\partial x^\nu} + \frac{\partial h_{\nu}^{.\:0}}{\partial x^\mu} - \frac{\partial h_{\mu\nu}}{\partial x_0}\bigg) = \frac{1}{2}\bigg(0 + 0 - 0 \bigg) = 0 \nonumber
    \end{equation}
thus the equation for photon trajectory is simple:
    \begin{equation}
        \frac{d^2 x^i}{dt^2} = -  \Gamma^i_{\mu\nu} \frac{d x^{\mu}}{dt} \frac{d x^{\nu}}{dt}, \;i = 1, 2, 3 \nonumber
    \end{equation}
The complete list of all Christoffel symbols is shown in the tables \ref{tab:cr1}, \ref{tab:cr2}, \ref{tab:cr3}:
    \begin{table}[h!]
        \centering
        \renewcommand{\arraystretch}{3}
        \begin{tabular}{c|c|c|c|c|}
             & $\nu = 0$ &$\nu = 1$ & $\nu = 2$ & $\nu = 3$ \\
            \hline
            $\mu = 0$ & 0 & 0 & 0 & 0 \\ \hline
            $\mu = 1$ & 0 & $-\dfrac{1}{2}\:\dfrac{\partial h_{11}}{\partial x^1}$ & $-\dfrac{1}{2}\:\dfrac{\partial h_{11}}{\partial x^2}$ & $-\dfrac{1}{2}\:\dfrac{\partial h_{11}}{\partial x^3}$ \\ \hline
            $\mu = 2$ & 0 & $-\dfrac{1}{2}\:\dfrac{\partial h_{11}}{\partial x^2}$ & $-\dfrac{\partial h_{21}}{\partial x^2}+\dfrac{1}{2}\:\dfrac{\partial h_{22}}{\partial x^1}$  & $-\dfrac{1}{2}\:\dfrac{\partial h_{12}}{\partial x^3}$  \\ \hline
            $\mu = 3$ & 0 & $-\dfrac{1}{2}\:\dfrac{\partial h_{11}}{\partial x^3}$  & $-\dfrac{1}{2}\:\dfrac{\partial h_{12}}{\partial x^3}$  & $\dfrac{1}{2}\:\dfrac{\partial h_{33}}{\partial x^1}$ \\ \hline
        \end{tabular}
        \caption{Evaluation of all $\Gamma^1_{\mu\nu}$}
        \label{tab:cr1}
    \end{table}

     \begin{table}[h!]
        \centering
        \renewcommand{\arraystretch}{3}
        \begin{tabular}{c|c|c|c|c|}
             & $\nu = 0$ &$\nu = 1$ & $\nu = 2$ & $\nu = 3$ \\
            \hline
            $\mu = 0$ & 0 & 0 & 0 & 0 \\ \hline
            $\mu = 1$ & 0 & $-\dfrac{\partial h_{12}}{\partial x^1}+\dfrac{1}{2}\:\dfrac{\partial h_{11}}{\partial x^2}$ & $-\dfrac{1}{2}\:\dfrac{\partial h_{22}}{\partial x^1}$ & $-\dfrac{1}{2}\:\dfrac{\partial h_{12}}{\partial x^3}$ \\ \hline
            $\mu = 2$ & 0 & $-\dfrac{1}{2}\:\dfrac{\partial h_{22}}{\partial x^1}$ & $-\dfrac{1}{2}\:\dfrac{\partial h_{22}}{\partial x^2}$  & $-\dfrac{1}{2}\:\dfrac{\partial h_{22}}{\partial x^3}$  \\ \hline
            $\mu = 3$ & 0 & $-\dfrac{1}{2}\:\dfrac{\partial h_{12}}{\partial x^3}$  & $-\dfrac{1}{2}\:\dfrac{\partial h_{22}}{\partial x^3}$  & $\dfrac{1}{2}\:\dfrac{\partial h_{33}}{\partial x^1}$ \\ \hline
        \end{tabular}
        \caption{Evaluation of all $\Gamma^2_{\mu\nu}$}
        \label{tab:cr2}
    \end{table}

    \newpage
    \begin{table}[h!]
        \centering
        \renewcommand{\arraystretch}{3}
        \begin{tabular}{c|c|c|c|c|}
             & $\nu = 0$ &$\nu = 1$ & $\nu = 2$ & $\nu = 3$ \\
            \hline
            $\mu = 0$ & 0 & 0 & 0 & 0 \\ \hline
            $\mu = 1$ & 0 & $\dfrac{1}{2}\:\dfrac{\partial h_{11}}{\partial x^3}$ & $\dfrac{1}{2}\:\dfrac{\partial h_{12}}{\partial x^3}$ & $-\dfrac{1}{2}\:\dfrac{\partial h_{33}}{\partial x^1}$ \\ \hline
            $\mu = 2$ & 0 & $\dfrac{1}{2}\:\dfrac{\partial h_{12}}{\partial x^3}$ & $\dfrac{1}{2}\:\dfrac{\partial h_{22}}{\partial x^3}$  & $-\dfrac{1}{2}\:\dfrac{\partial h_{33}}{\partial x^2}$  \\ \hline
            $\mu = 3$ & 0 & $-\dfrac{1}{2}\:\dfrac{\partial h_{33}}{\partial x^1}$ & $-\dfrac{1}{2}\:\dfrac{\partial h_{33}}{\partial x^2}$ & $-\dfrac{1}{2}\:\dfrac{\partial h_{33}}{\partial x^3}$ \\ \hline
        \end{tabular}
        \caption{Evaluation of all $\Gamma^3_{\mu\nu}$}
        \label{tab:cr3}
    \end{table}
    
    Recall $v_i = dx^i / dt$ and the equations are:

    \begin{equation}
        \begin{cases}
            \dfrac{d v^1}{dt} = \dfrac{1}{2}\:\dfrac{\partial h_{11}}{\partial x^1} \: \big( v^1 \big)^2 +\bigg(\dfrac{\partial h_{21}}{\partial x^2}-\dfrac{1}{2}\:\dfrac{\partial h_{22}}{\partial x^1}\bigg) \big( v^2 \big)^2 - \dfrac{1}{2}\:\dfrac{\partial h_{33}}{\partial x^1} \big( v^3 \big)^2 + 
            \\
            \\
            + \dfrac{\partial h_{11}}{\partial x^2} v^1 v^2 + \dfrac{\partial h_{11}}{\partial x^3} v^1 v^3 + \dfrac{\partial h_{12}}{\partial x^3} v^2 v^3 
            \\
            \\
            \dfrac{d v^2}{dt} = \dfrac{1}{2}\:\dfrac{\partial h_{22}}{\partial x^2} \: \big( v^2 \big)^2 +\bigg(\dfrac{\partial h_{21}}{\partial x^1}-\dfrac{1}{2}\:\dfrac{\partial h_{11}}{\partial x^2}\bigg) \big( v^1 \big)^2 - \dfrac{1}{2}\:\dfrac{\partial h_{33}}{\partial x^2} \big( v^3 \big)^2 + 
            \\
            \\
            + \dfrac{\partial h_{22}}{\partial x^1} v^1 v^2 + \dfrac{\partial h_{12}}{\partial x^3} v^1 v^3 + \dfrac{\partial h_{22}}{\partial x^3} v^2 v^3 
            \\
            \\
            \dfrac{d v^3}{dt} = - \dfrac{1}{2}\:\dfrac{\partial h_{33}}{\partial x^3}\big( v^3 \big)^2 - \dfrac{1}{2}\:\dfrac{\partial h_{11}}{\partial x^3}\big( v^1 \big)^2 - \dfrac{1}{2}\:\dfrac{\partial h_{22}}{\partial x^3}\big( v^2 \big)^2 -
            \\
            \\
            - \dfrac{1}{2}\:\dfrac{\partial h_{33}}{\partial x^1} v^1 v^3
            - \dfrac{1}{2}\:\dfrac{\partial h_{33}}{\partial x^2} v^2 v^3
            - \dfrac{1}{2}\:\dfrac{\partial h_{12}}{\partial x^3} v^1 v^2
        \end{cases} \nonumber
    \end{equation}

    The lensing effect is small, so we can just adopt the first order expansion in $
    \mu$ for $v^1$ and $v^2$. We can use an approximation $v^3 = 1$ since it will be the second order correction at most in the first two equations:
    \begin{equation}\label{1stordertraj}
        \begin{cases}
            \dfrac{d v^1}{dt} = - \dfrac{1}{2}\:\dfrac{\partial h_{33}}{\partial x^1} + \dfrac{\partial h_{11}}{\partial x^3} v^1 + \dfrac{\partial h_{12}}{\partial x^3} v^2
            \\
            \\
            \dfrac{d v^2}{dt} = - \dfrac{1}{2}\:\dfrac{\partial h_{33}}{\partial x^2} + \dfrac{\partial h_{12}}{\partial x^3} v^1 + \dfrac{\partial h_{22}}{\partial x^3} v^2
            \\
            \\
            v^3 = 1
        \end{cases} 
    \end{equation}

    Using the third equation in (\ref{1stordertraj}), we can replace the $d/dt$ by $d/dz$ and this is the final result before the derivation of lens equation.

\section{Solving a lens equation for a curved CS}\label{secA4}

Before the derivation we need to solve the linearized Einstein's equation for a static bended string:
\begin{equation}
    h_{\mu\nu}(\mathbf{x}) = 4G \int_{\mathbb{R}^3} d\mathbf{x'}\:\frac{S_{\mu\nu}}{|\mathbf{x} - \mathbf{x'}|} \nonumber
\end{equation}

For the purpose of simplicity we will consider the case $R \ll R_s \Delta \theta $ so the field from bend $s\in [-R, R]$ can be neglected. Recalling:
\begin{equation}
    h_{\uparrow,\downarrow} = \frac{\Delta \theta}{2\pi}\int_{\mathbb{R}^3} d\mathbf{x'}\:\frac{\delta_{\uparrow, \downarrow}}{|\mathbf{x} - \mathbf{x'}|} \nonumber
\end{equation}
we can rewrite geodesic equations (\ref{1stordertraj}):
\begin{equation}\label{geodesicsString}
    \frac{d \mathbf{v}}{dz} =  -\underbrace{ \frac{1}{2}
    \left(
    \begin{array}{c}
        \partial_x \\
        \partial_y
    \end{array}
    \right)
    (h_\uparrow + h_\downarrow) }_{=\mathbf{A}(x, y, z)} -  \underbrace{\bigg[\frac{\partial h_\uparrow}{\partial z}
    \left(
    \begin{array}{cc}
        \sin^2\theta & \sin\theta\cos\theta \\
        \sin\theta\cos\theta & \cos^2\theta
    \end{array}
    \right)
    + \frac{\partial h_\downarrow}{\partial z}
    \left(
    \begin{array}{cc}
        0 & 0 \\
        0 & 1
    \end{array}
    \right)\bigg]}_{=\mathbf{\hat B}(x, y, z)}
    \mathbf{v} 
\end{equation}
where $\mathbf{v} = (v^1, v^2)^T$. This equation is particularly important since in linear approximation $\mathbf{v}$ is an angle between the photon path and the $z$-axis. Suppose that without the string lensing a point is seen from direction $\mathbf{n}$,  $\mathbf{v}_i$ is the initial direction of photon's path and $-\mathbf{v}_f$ is the final direction under which the point is seen with a lens. Thus, the lens equation is:
\begin{equation}
    \mathbf{n} = -\mathbf{v}_f - (\mathbf{v}_f - \mathbf{v}_i)\frac{R_g - R_s}{R_g} \nonumber
\end{equation}
If we know the initial image without the lens, then we know $\mathbf{n}$. Our task is to find $\mathbf{v}_f$. The connection between $\mathbf{v}_i$ and $\mathbf{v}_f$ is presented by the solution of DE (\ref{geodesicsString}) with the initial condition $\mathbf{v}(R_g \mathbf{n}, z = 0) = -\mathbf{v}_i$. This approximation is valid only in the case, when all the lensing happens at $z = R_g - R_s$, near the string. But the string is not a compact object, so we propose another scheme.

Suppose the radiation is emitted from the point $\mathbf{r}(z = 0) = R_g \mathbf{n}_0$. $\mathbf{r} = (x(z), y(z))^T$ is a vector in the picture plane with a fixed value of $z$. The ray should have the initial conditions $\mathbf{v_0}$ such that it travels to the observer, so $\mathbf{r}(z = R_g) =0$. Thus the boundary value problem can be formulated:
\begin{equation}\label{IBP}
    \begin{cases}
        d\mathbf{v}/dz = -\mathbf{A}(\mathbf{r}, z) - \mathbf{\hat B}(\mathbf{r}, z)\mathbf{v}
        \\
        
        d\mathbf{r}/dz = \mathbf{v}
        
        \\
        \mathbf{r}(z = 0) = R_g \mathbf{n}_0
        \\
        \mathbf{r}(z = R_g) =0
    \end{cases} 
\end{equation}
Once this problen is solved, we can calculate $\mathbf{v}(z = R_g) = -\mathbf{v}_f$ and create a map $\mathbf{v}_f(\mathbf{n}_0)$, which is our lens equation. 

To solve this equation numerically, we can treat $\mathbf{v}_f$ as a parameter in shooting method for IBP (\ref{IBP}). To further simplify the process of numerical integration, we rescale the spatial variables:
\begin{equation}
    \begin{cases}
        a = z / R_g \in [0, 1]\\
        \mathbf{n} = \mathbf{r} / R_g
    \end{cases} \nonumber
\end{equation}
and the IBP reads:
\begin{equation}
    \begin{cases}
        \dfrac{d \mathbf{v}}{da} =  -\dfrac{1}{2}
        \nabla_\mathbf{n}
        (h_\uparrow + h_\downarrow) -  
        \bigg[\dfrac{\partial h_\uparrow}{\partial a}
        \left(
        \begin{array}{cc}
            \cos^2\theta & \sin\theta\cos\theta \\
            \sin\theta\cos\theta & \sin^2\theta
        \end{array}
        \right)
        + \dfrac{\partial h_\downarrow}{\partial a}
        \left(
        \begin{array}{cc}
            1 & 0 \\
            0 & 0
        \end{array}
        \right)\bigg]
        \mathbf{v}\\
        
        d\mathbf{n}/da = \mathbf{v}
        
        \\
        \mathbf{n}(a = 0) = \mathbf{n}_0
        \\
        \mathbf{n}(a = 1) =0
    \end{cases} \nonumber
\end{equation}

We also need all the derivatives of string metric: $\nabla_{\mathbf{n}} h_{\uparrow, \downarrow}$, $\partial h_{\uparrow, \downarrow} / \partial a$. They are (in terms of scaled variables, $r = R_s / R_g$): 
\begin{equation}
    \frac{\partial h_\uparrow}{\partial a} = \frac{\Delta \theta}{2\pi} \: \frac{a - 1 + r}{(a - 1 + r)^2 + (n_x \cos\theta - n_y \sin\theta)^2}\bigg( 1 + \frac{n_x\sin\theta + n_y \cos\theta}{\sqrt{\mathbf{n}^2 + (a - 1 + r)^2}} \bigg) \nonumber
\end{equation}
\begin{equation}
    \frac{\partial h_\downarrow}{\partial a} = \frac{\Delta \theta}{2\pi} \: \frac{a - 1 + r}{(a - 1 + r)^2 + n_x^2}\bigg( 1 - \frac{n_y}{\sqrt{\mathbf{n}^2 + (a - 1 + r)^2}} \bigg) \nonumber
\end{equation}
\begin{equation}
    \frac{\partial h_\uparrow}{\partial n_x} = \frac{\Delta \theta}{2\pi} \: \bigg[-\frac{\sin\theta}{\sqrt{\mathbf{n}^2 + (a - 1 + r)^2}} + \nonumber
\end{equation}
\[
    + \cos\theta \frac{n_x \cos\theta - n_y \sin\theta}{(a - 1 + r)^2 + (n_x \cos\theta - n_y \sin\theta)^2}\bigg( 1 + \frac{n_x\sin\theta + n_y \cos\theta}{\sqrt{\mathbf{n}^2 + (a - 1 + r)^2}}  \bigg)\bigg]
\]
\begin{equation}
    \frac{\partial h_\downarrow}{\partial n_x} = \frac{\Delta \theta}{2\pi} \: \frac{n_x}{(a - 1 + r)^2 + n_x^2}\bigg( 1 - \frac{n_y}{\sqrt{\mathbf{n}^2 + (a - 1 + r)^2}} \bigg) \nonumber
\end{equation}
\begin{equation}
    \frac{\partial h_\uparrow}{\partial n_y} = \frac{\Delta \theta}{2\pi} \: \bigg[-\frac{\cos\theta}{\sqrt{\mathbf{n}^2 + (a - 1 + r)^2}} - \nonumber
\end{equation}
\[
    - \sin\theta \frac{n_x \cos\theta - n_y \sin\theta}{(a - 1 + r)^2 + (n_x \cos\theta - n_y \sin\theta)^2}\bigg(1 + \frac{n_x\sin\theta + n_y \cos\theta}{\sqrt{\mathbf{n}^2 + (a - 1 + r)^2}} \bigg)\bigg]
\] 
\begin{equation}
    \frac{\partial h_\downarrow}{\partial n_y} = \frac{\Delta \theta}{2\pi} \: \frac{1}{\sqrt{\mathbf{n}^2 + (a - 1 + r)^2}} \nonumber
\end{equation}
\end{appendices}

%%%%%%%%%%%%%%%%%%%%%%%%%%%%%%%%%%%%%
%\bibliography{sn-bibliography}
\bibliography{EPJC_new/sn-article}{}

\end{document}